\documentclass[aps,showpacs,pre,twocolumn,superscriptaddress]{revtex4}
\usepackage{amsfonts}
\usepackage{amsmath}
\usepackage{mathrsfs}
\usepackage{amssymb}
\usepackage{wasysym}
\usepackage{subfigure}
\usepackage{graphicx}
\providecommand{\U}[1]{\protect\rule{.1in}{.1in}}

\begin{document}
\title{Quantum Phase diagram and time-of-flight absorption pictures of ultracold Bose system in a square optical superlattice}
\author{Fan Wei}
\affiliation{Department of Physics, Shanghai University, Shanghai 200444, P.R. China}
\author{Jun Zhang}
\affiliation{Department of Physics, Shanghai University, Shanghai 200444, P.R. China}
\author{Ying Jiang}
\thanks{Corresponding author}
\email{yjiang@shu.edu.cn}
\affiliation{Department of Physics, Shanghai University, Shanghai 200444, P.R. China}
\affiliation{Qian Weichang College, Shanghai University, Shanghai 200444, P.R. China}

\begin{abstract}
In this letter, by the use of the generalized effective potential theory, with the help of process-chain approach under the framework of Kato formulation of perturbation expansion, we calculate out the quantum phase diagram up to 8-th order for an ultracold Bose system in a square optical superlattice. Base on these perturbative data, with the help of the linear fit extrapolation technique, more accurate results are gotten, which are in excellent agreement with recent Monte-Carlo numerical results. Moreover, by employing the generalized re-summed Green's function method and cumulant expansion, the momentum distribution function of the system is also calculated analytically and the time-of-flight absorption pictures of the system are plotted.
\end{abstract}


\maketitle

- {\it Introduction.} - The physics of ultracold atomic systems in optical lattices has been boosted as one of the major and fascinating fields during past decade \cite{Greiner,Lewenstein-book}, and has received tremendous attention on account of the novelty and various potential applications\cite{bloch,bloch-2014}, ranging from condensed matter physics \cite{capogrosso2,zhouqi}, quantum information processing \cite{Jaksch-2,englund,hazzard}, even to cosmological problems \cite{steinhauer}.

While many experiments with ultracold atoms have been performed in homogeneous simple cubic lattices, more and more experimental and theoretical efforts have been devoted to complex systems, including ultracold bosons in non-rectangular optical lattices \cite{becker,soltan-panahi,jo}, systems with long range interactions \cite{pfau,hild}, or with multi components\cite{soltan}. Among them, an interesting topic is ultracold Bose systems in optical superlattices \cite{folling,cheinet}. Due to the complexity of the lattice structures, the quantum phase transitions and corresponding phase diagrams of these systems are expected to be more complex, and hence need to be investigated analytically.

Actually, beside the mean-field theory \cite{fisher}, which underestimates the phase boundaries when compared with Monte-Carlo results \cite{capogrosso1}, and the strong coupling expansion method \cite{freericks1}, which overestimates the phase boundaries, an alternative field theoretical method, the so-called effective potential theory  \cite{santos} has been developed in recent year. By utilizing it, the quantum phase diagrams of ultracold Bose systems in triangular, hexagonal, and Kagom\'e optical lattices have been investigated \cite{jiang-pra-1} , the relative difference between our analytical results and numerical results \cite{elstner,teichmann2} is less than $10\%$.

Meanwhile, momentum distribution functions, which can be revealed via time-of-flight technique \cite{bloch,Greiner,kohl,becker}, should be calculated as well in order to compare the theoretical results with experimental data. Thanks to re-summed Green's function method \cite{ohliger} and cumulant expansion \cite{metzner1}, we have calculated out analytically the time-of-flight absorption pictures for ultracold Bose systems in triangular optical lattices for various lattice depth \cite{jiang-pra-2}, exhibiting good agreement with experimental data \cite{becker}.

In this letter, we will investigate analytically the quantum phase transitions of a ultracold Bose system in a square superlattice. In fact, Wang {\it et al.} \cite{wangtao} have developed a generalized effective potential theory and calculated out the phase diagram of an ultracold superlattice Bose system up to second order. With the help of process chain technique \cite{teichmann}, by the use of Kato formulation \cite{kato} of perturbation, we will push the limit of the perturbative calculation of quantum phase boundaries of such superlattice ultracold Bose system up to 8th order, and further goes to the order of infinity via the technique of linear fit extrapolation. Moreover, by making use of the Green's function method, we will calculate the time-of-flight absorption pictures of the superlattice ultracold Bose system.

- {\it The generalized effective potential theory.} -
A system of spinless bosons trapped in a square superlattice is described by\cite{buonsante,wangtao}
\begin{eqnarray}
\label{H_SL}
\hat{H}_{SL}=&&-t\underset{\left\langle i\in A,j\in B\right\rangle }{\sum}\left(\hat{a}_{i}^{\dagger}\hat{a}_{j}+\hat{a}_{j}^{\dagger}\hat{a}_{i}\right)
+\frac{U}{2}\underset{i\in A,B}{\sum}\hat{n}_{i}\left(\hat{n}_{i}-1\right) \nonumber \\
&&-\left(\mu+\Delta\mu\right)\underset{i\in A}{\sum}\hat{n}_{i}-\mu\underset{i\in B}{\sum}\hat{n}_{i}
\end{eqnarray}
where $t$ is the nearest-neighbor hopping parameter, $U$ denotes the on-site repulsion between two atoms. $\hat{a}_{i}^{\dagger}\left(\hat{a}_{i}\right)$ represents the boson creation(annihilation) operator at site $i$, $\Delta \mu$ stands for the difference of the chemical potentials on sublattice A and B, without loss of generality, we set $\Delta \mu>0$.

The subtle balance among the hopping parameter $t$, on-site repulsion $U$, and $\Delta \mu$ introduces rich phase structures to this system. Even when the hopping parameter $t=0$, the competition between $U$ and $\Delta \mu$ will lead to different quantum phases at zero temperature, these phases correspond to different regions of the chemical potential.

Before we discuss the quantum phase boundaries of the system, the ground state of the site-diagonal part of the above Hamiltonian
\begin{equation}
\label{H0_AB}
\hat{H}_{0}=\frac{U}{2}\sum_{i\in A,B}\hat{n}_{i}\left(\hat{n}_{i}-1\right)-(\mu+\Delta \mu)\sum_{i\in A}\hat{n}_{i}-\mu\sum_{i\in B}\hat{n}_{i}
\end{equation}
need to be determined.
Since all terms in this part of Hamiltonian are site-diagonal, its eigenstates can be denoted in terms of the occupation number of each sublattice, i.e. $|n_A,n_B\rangle$, the corresponding eigenenergies read
\begin{eqnarray}
&E_{n_{A},n_{B}}&=\frac{U}{2}n_{A}(n_{A}-1)+\frac{U}{2}n_{B}(n_{B}-1)\nonumber \\
&&-\mu(n_{A}+n_{B})-\Delta \mu n_A.
\label{eigenvalue}
\end{eqnarray}
Mott insulator phases correspond to the cases when $n_A=n_B$, while charge-density-wave (CDW) phases being $n_A\neq n_B$. As is known, the condition for the existence of Mott phase $|n,n\rangle$ is $E_{n,n}\leq E_{n+1,n}$ and at the same time $E_{n,n}\leq E_{n,n-1}$, from Eq.(\ref{eigenvalue}), we see that this leads to
\begin{equation}
U(n-1)\leq \mu \leq Un -\Delta \mu.
\end{equation}
It is easy to recognize that there is no Mott phase when $\Delta \mu>U$, and the ground state for $t=0$ at any case would be CDW phases. Hence, without loss of generality, we would focus on the case of $0<\Delta \mu <U$ in the rest of the paper. In this case, the occupation number of these two sublattices in possible CDW phases may only be $n_B=n_A-1$, and the corresponding regions of $\mu$ read
\begin{equation}
U(n-1)-\Delta \mu\leq \mu \leq U(n-1).
\end{equation}

With this information in hand, we can now go further to tune on the hopping parameter to see the quantum phase transitions from the incompressible Mott states or CDW states to superfluid phase. In order to investigate the quantum phase transitions of the superlattice Bose system and determine the corresponding phase boundaries analytically, similar to previous work \cite{santos,jiang-pra-1}, we would like to use the generalized effective potential method \cite{wangtao} under the framework of Ginzburg-Landau field theory, i.e. we add, for the moment, additional source terms with strength $\bold{J}=(J_A, J_B)^T$ into the above superlattice Bose-Hubbard Hamiltonian as following
\begin{equation}
\label{H_AB}
\hat{H}_{SL}\!\!\left({\bold J},{\bold J}^{\dagger}\right)\!\!=\!\!\hat{H}_{SL}\!\!+\!\!\underset{j\epsilon A}{\sum}\!\!\left(\!J_{A}\hat{a}_{j}^{\dagger}\!+\!J_{A}^{*}\hat{a}_{j}\!\right)\!\!+\!\!\underset{j\epsilon B}{\sum}\!\!\left(\!J_{B}\hat{a}_{j}^{\dagger}\!+\!J_{B}^{*}\hat{a}_{j}\!\right)
\end{equation}
and treat the hopping term and the additional external source terms as perturbations.

It is quite straightforward to get the grand canonical free energy of the system as power series of both the hopping parameter $t$ and the external source ${\bold J}$ (${\bold J}^{\dagger}$) via Taylor expansion. Since the unperturbed ground states are either Mott phases or CDW states, ${\bold J}$ and ${\bold J}^{\dagger}$ can only appear in pair, after re-grouping the terms in the free energy with respect to ${\bold J}$ and ${\bold J}^{\dagger}$, the free energy reads
\begin{equation}
\label{FreeEnergy_sl}
F\!\left({\bold J},\!{\bold J}^{\dagger}\!,t\right)\!=\!N_{s}\!\left[\!F_{0}\left(t\right)\!+\!{\bold J}^{\dagger}\!{\mathbb C}_{2}\!\left(t\right)\!{\bold J}\!+\!{\bold J}^{\dagger}\!{\bold J}^{\dagger}\!{\mathbb C}_{4}\!\left(t\right)\!{\bold J}{\bold J}\!+\!\cdots\!\right]
\end{equation}
with the expansion coefficient tensor ${\mathbb C}_2(t)=\left(\begin{array}{cc}
c_{2AA} & c_{2AB}\\
c_{2BA} & c_{2BB}
\end{array}\right)$ being $
c_{2AA}=\sum_{n=0}^{\infty}\left(-t\right)^{n}\alpha_{2AA}^{\left(n\right)}$,
$c_{2BB}=\sum_{n=0}^{\infty}\left(-t\right)^{n}\alpha_{2BB}^{\left(n\right)}$,
$c_{2AB}=\sum_{n=0}^{\infty}\left(-t\right)^{n}\alpha_{2AB}^{\left(n\right)},
$
$N_s$ is the total number of the lattice sites.

Due to the nature of the superlattice structure of the system, the superfluid order parameter takes the form of vector ${\bold \Psi}=(\psi_A, \psi_B)^T$ (${\bold \Psi}^{\dagger}=(\psi_A^*, \psi_B^*)$) whose components are defined as $\psi_m=\langle a_m\rangle$ ($\psi_m^*=\langle a_m^{\dagger}\rangle$) ($m=A, B$)
which can be calculated from the free energy by \cite{kleinert-phi4}
$\psi_{m}=\frac{1}{N_{S}}\frac{\partial F}{\partial J_{m}^{*}}$ ($
\psi_{m}^{*}=\frac{1}{N_{S}}\frac{\partial F}{\partial J_{m}}$) ($m=A,B
$).

A Legendre transformation of the free energy $F\left({\bold J},{\bold J}^{\dagger},t\right)$ leads to an effective potential $\Gamma\left({\bf \Psi},{\bf \Psi}^{\dagger},t\right)=F/N_{s}-{\bold \Psi}^{\dagger}{\bold J}-{\bold J}^{\dagger} {\bold \Psi}$
who is expressed in terms of ${\bf \Psi}$ (${\bf \Psi}^{\dagger}$) as
\begin{equation}
\label{Landau}
\Gamma\!\left(\!{\bold \Psi}\!,\!{\bold \Psi}^{\dagger}\!,\!t\!\right)\!=\!F_{0}\left(t\right)\!+\!{\bold \Psi}^{\dagger}\!{\mathbb A}_{2}\!\left(t\right)\!{\bold \Psi}\!+\!{\bold \Psi}^{\dagger}\!{\bold \Psi}^{\dagger}\!{\mathbb A}_{4}\!\left(t\right)\!{\bold \Psi}\!{\bold \Psi}\!+\!\cdots,
\end{equation}
this takes exactly the form of Landau $\phi^4$ theory. According to the Landau theory, $\det {\mathbb A}_{2}\left(t\right)=0$ determines the phase boundaries between the uncompressed Mott or CDW states and superfluid phases. It is not difficult to find out that ${\mathbb A}_{2}\left(t\right)=-{\mathbb C}_{2}^{-1}\left(t\right)$, hence, the critical value of $t$ can be found by looking for the radius of convergence of the determinant of ${\mathbb C}_{2}\left(t\right)$
\begin{equation}
\det {\mathbb C}_{2}\left(t\right)\!=\!c_{2AA}c_{2BB}\!-\!c_{2AB}c_{2BA}\equiv\!\sum_{k=0}^{\infty}\!D_2^{(2k)}t^{2k} \end{equation}
There is no odd order term of $t$ because that $\alpha_{2AA\left(BB\right)}$ only has even terms while $\alpha_{2AB\left(BA\right)}$ only has odd terms.
The convergence radius may be found via the so-called d'Alembert's ratio test, and the $n$-th order approximation of the phase boundary reads
\begin{widetext}
\begin{equation}
t^{2}_c=\frac{D_2^{(n-2)}}{D_2^{(n)}}=\frac{\alpha_{2AA}^{\left(n-2\right)}\alpha_{2BB}^{\left(0\right)}-\alpha_{2AB}^{\left(n-3\right)}\alpha_{2BA}^{\left(1\right)}+
\cdots-\alpha_{2AB}^{\left(1\right)}\alpha_{2BA}^{\left(n-3\right)}+\alpha_{2AA}^{\left(0\right)}\alpha_{2BB}^{\left(n-2\right)}}{\alpha_{2AA}^{\left(n\right)}
\alpha_{2BB}^{\left(0\right)}-\alpha_{2AB}^{\left(n-1\right)}\alpha_{2BA}^{\left(1\right)}+\cdots-\alpha_{2AB}^{\left(1\right)}\alpha_{2BA}^{\left(n-1\right)}
+\alpha_{2AA}^{\left(0\right)}\alpha_{2BB}^{\left(n\right)}}.
\label{t}
\end{equation}
\end{widetext}

\begin{figure}[here]
\includegraphics[width=0.6\linewidth]{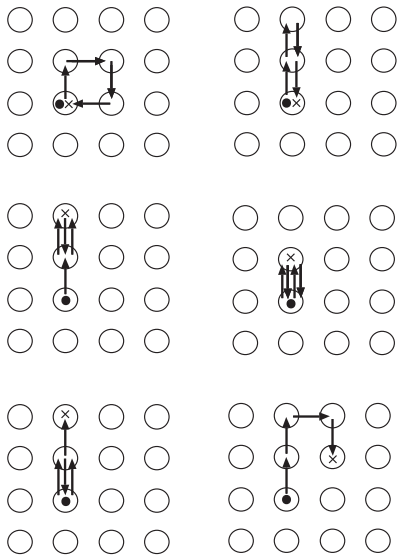}
\caption{Sketch diagrams for topologically different processes of sixth order perturbation terms, consisting of exactly one creation ($\bullet$), one annihilation ($\times$) and four nearest-neighboring hoppings (arrows in diagrams).}
\label{sixorder}
\end{figure}

- {\it Process chain calculation and the quantum phase diagram.} -
From the above discussion, we see that in order to find the critical value of $t$ to determine the phase boundaries, the coefficients of $\alpha^{(n)}_{2}$ have to be calculated out, this can only be achieved perturbatively. Since the unperturbed state is Mott state or CDW state, every nonzero contribution to $\alpha^{(n)}_{2}$ includes exactly one creation operator at site-$i$ (associated with $J_i$) and one annihilation operator at site-$j$ (associated with $J^*_j$) as well as $n$ nearest-neighbor hopping processes (associated with $t$) connecting site-$i$ and $j$.

Actually, when calculating many-body problems, perturbation theory as Rayleigh-Schrodinger perturbation expansion\cite{gelfand} may be applied, with the help of Kato's formulation of perturbation series \cite{kato} and process chain technique \cite{teichmann}, results with very high order correction may in principle be reached. Let us denote the ground state of $\hat{H}_0$ by $\left|m\right\rangle$. In general, when $|m\rangle$ is subject to some perturbation $V$, the eigenenergy of the total Hamiltonian $\hat{H}=\hat{H}_0+V$ can be perturbatively expressed as $E_m=E_m^{(0)}+\sum_nE_m^{(n)}$ with the $n$-th order correction given by the trace\cite{kato}
\begin{equation}
\label{trace}
E_{m}^{\left(n\right)}={\rm Tr}\left[{\displaystyle \sum_{\left\{ \alpha_{\ell}\right\} }S^{\alpha_{1}}V}S^{\alpha_{2}}VS^{\alpha_{3}}\ldots S^{\alpha_{n}}VS^{\alpha_{n+1}}\right],
\end{equation}
the operators $S^{\alpha}$ connecting each perturbation terms $V$ read
\begin{equation}
\label{op_S}
S^{\alpha}=\left\{ \begin{array}{cc}
-\left|m\right\rangle \left\langle m\right| & for\;\alpha=0\\
{\displaystyle \sum_{i\neq m}\frac{\left|i\right\rangle \left\langle i\right|}{\left(E_{m}^{\left(0\right)}-E_{i}^{\left(0\right)}\right)^{\alpha}}} & for\;\alpha>0
\end{array}\right.
\end{equation}
with $E^{(0)}_{m}$ and $E^{(0)}_{i}$ denote the unperturbed energies of the $\hat{H}_{0}$'s eigenstates  $\left|m\right\rangle$ and $\left|i\right\rangle$. Since the eigenstates of $\hat{H}_0$ form an orthonormal basis, it is easily to be proved that $
S^{0}S^{\alpha}=0,\quad {\rm for} \quad \alpha > 0$, and $
S^{\alpha}S^{\beta}=S^{\alpha+\beta},\quad {\rm for} \quad \alpha,\beta > 0$.
It should be emphasized that all possible sequences of $\left\{ \alpha_{\ell}\right\}$ with constraint $\sum\limits_{\ell=1}^{n+1}\alpha_{\ell}=n-1$ must be taken into account in the calculation \cite{kato,teichmann}.

Each Kato term $\langle m|S^{\alpha_{1}}V S^{\alpha_{2}}VS^{\alpha_{3}}\ldots S^{\alpha_{n}}VS^{\alpha_{n+1}}|m\rangle$ in Eq.(\ref{trace}) represents a sum of processes going from the ground state $\left|m\right\rangle$ over series of different intermediate states $\left|i\right\rangle$, caused by the perturbation $V$, then back to $\left|m\right\rangle$, i.e. the Kato terms are sums of so-called process chains \cite{teichmann,teichmann2}.

In our case, the perturbation terms are hopping terms and external source terms,
and the unperturbed ground state $|m\rangle$ stands for the Mott or CDW state. All these process chains can be abstractly represented by diagrams \cite{teichmann}, due to the linked-cluster theorem \cite{gelfand}, only connected diagrams contribute. As an example, we show in Fig. \ref{sixorder} topologically different diagrams for sixth order perturbations consisting of exactly one source in (denoted by a dot $\left(\bullet\right)$ in the diagram), one source out (denoted by $\left(\times\right)$) and four hoppings between nearest-neighbor sites (denoted by arrows).

\begin{figure}[here]
\includegraphics[width=0.9\linewidth]{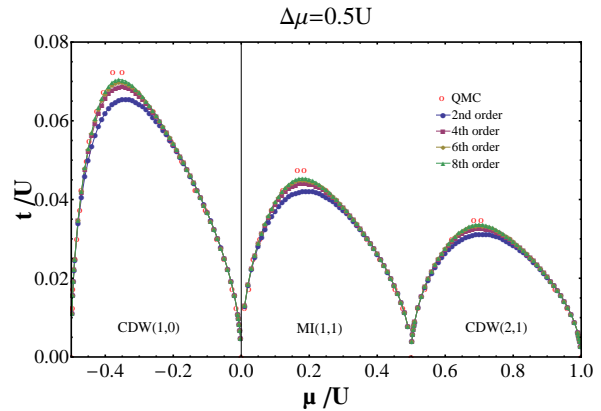}
\caption{(Color online)The quantum phase diagram of ultra-cold Bose system in a square superlattice with $\Delta \mu=0.5U$. The quantum Monte-Carlo numerical simulation result \cite{wangtao} is also shown. }
\label{phase-boundary}
\end{figure}

By applying the above sketch process chain method to the calculation of all $\alpha^{(n)}_2$ needed, together with Eq. (\ref{t}), the phase boundaries between Mott (or CDW) state and superfluid phase can then be calculated. The quantum phase boundaries of the ultra-cold Bose system in a square superlattice with $\Delta \mu=0.5U$ are shown in Fig. \ref{phase-boundary} up to 8-th order. The comparison to Monte-Carlo numerical results \cite{wangtao} shows that the relative deviation of our 8-th order process chain results from the Monte-Carlo results is less than $4\%$.

\begin{figure}
\includegraphics[width=0.9\linewidth]{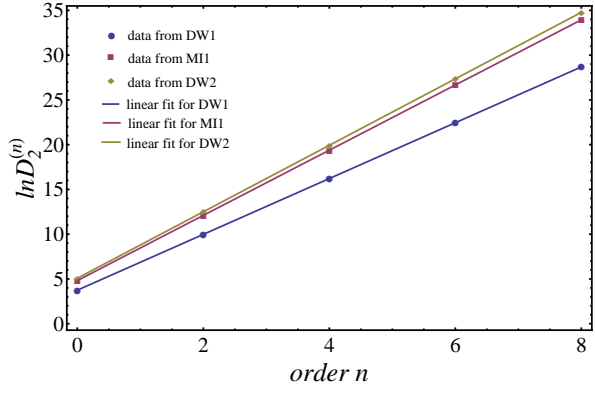}
\caption{(Color online) Logarithm of $D_2^{(n)}$ versus the order $n$ in different cases, lines are corresponding linear fits.}
\label{linearfit1}
\end{figure}

\begin{figure}
\includegraphics[width=0.9\linewidth]{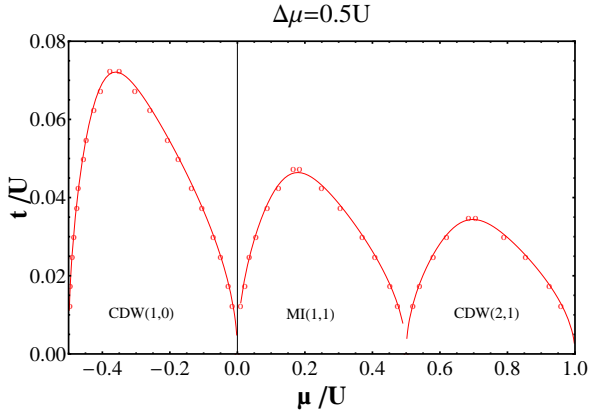}
\caption{(Color online) The quantum phase diagram of ultra-cold Bose system in a square superlattice with $\Delta \mu=0.5U$ from the extrapolation. The quantum Monte-Carlo numerical simulation result \cite{wangtao} is also shown. A good agreement is exhibited. }
\label{phase-boundary-extr}
\end{figure}

A more accurate result may further be gotten via linear fit extrapolation method \cite{teichmann2,teichmann}. When plotting the logarithm of the coefficients $D_2^{(n)}$ against the order of $n$, i.e. the number of tunneling processes in the perturbative calculation, for different ground states (CDW or MI), we find a good linear behavior, as shown in Fig. \ref{linearfit1}. This indicates that, in a good approximation, the ratio between the coefficients of adjacent orders is constant, and would be exact in the case of infinite dimensionality \cite{teichmann}. However, due to the tunneling perturbation, the ratio would change slightly when increasing the order $n$. As is known, in order to get the exact convergence radius of a power series via d'Alembert's ratio test, the order $n$ should be sent to infinite. Based on the data of order 2, 4, 6, and 8 gotten above, the extrapolation over $1/n$ can be carried out, and we can then obtain the critical value of $t_c$ for different $\mu$ when $n\rightarrow \infty$. As is shown in Fig. \ref{phase-boundary-extr}, our result is nearly identical with the numerical result \cite{wangtao}. By selecting different orders when doing the extrapolation, we find that the relative error of our result is less than 1\%.

- {\it The Time-of-flight pictures.} -
In our previous work \cite{jiang-pra-2}, with the help of re-summed Green's function method \cite{ohliger,jiang-pra-2}, we have calculated analytically the time-of-flight (TOF) pictures of a Bose-Hubbard system in a triangular optical lattice. A generalized Green's function method \cite{lin-arxiv} has also been developed to tackle problems of a Bose-Hubbard system  in a bipartite optical lattice. In this section, by utilizing the generalized Green's function method, we will investigate the TOF of the  Bose-Hubbard model on a square superlattice.

As is known, the superlattice Bose-Hubbard Hamiltonian in Eq.(\ref{H_SL}) is the single-band approximation of the many-body Hamiltonian $
H\!\!=\!\!\int d{\bf r}\psi^{\dagger}({\bf r})\!\!\left[\!\!-\frac{\hbar^2 \nabla^2}{2m}\!+\!V\!({\bf r})\!\!\right]\!\!\psi({\bf r})\!\!+\!\!\frac g2\!\int d{\bf r}\psi^{\dagger}({\bf r})\psi^{\dagger}({\bf r})\psi({\bf r})\psi({\bf r})
$
 in tight-binding limit at extremely low temperature \cite{Jaksch},
the optical lattice potential $V({\bf r})$ is \cite{paul}
\begin{eqnarray}
V_{op}({\bf r})&=&-V_{0}\Big[\cos^{2}\left(\frac{2\pi x}{\lambda}\right)+\cos^{2}\left(\frac{2\pi y}{\lambda}\right) \nonumber \\
&&+2\cos\theta \cos\left(\frac{2\pi x}{\lambda}\right)\cos\left(\frac{2\pi y}{\lambda}\right)\Big]
\label{lattice-potential-1}
\end{eqnarray}
who create the checkerboard pattern of the square optical superlattice with alternate deep and shallow wells,
the relative well depth can be tuned in real time by changing the phase difference $\theta$ ($0<\cos\theta <1$)between the counterpropagating laser beams.
The lattice constant $a=\frac{\lambda}{2}$ with $\lambda$ being the wave length of the laser.

In the harmonic approximation, we expand the optical lattice potential in the vicinity of sublattice $A$ (the locations of deep wells, the sublattice which the site $(0,0)$ belongs to) and sublattice $B$ (for instance the site $(a,0)$), respectively, and easily get the harmonic-oscillator frequencies $\omega_{A,B}=\sqrt{\frac{V_{0}2\pi^{2}\left(1\pm\cos\theta\right)}{ma^{2}}}$ for deep wells ($+$ sign) and shallow wells ($-$ sign). Correspondingly, in unit of recoil energy $E_R=\pi^2\hbar^2/(2ma^2)$, the lowest band Wannier functions for sublattice $A$ and $B$ read
$
w\left({\bf r}-{\bf r}_{i}\right)|_{i\in A,B}=\left[\tilde{V}_{0}\left(1\pm\cos\theta\right)\right]^{\frac{1}{4}}\left(\frac{\pi}{a^{2}}\right)^{\frac{1}{2}}
\exp\big[-\frac{\pi^{2}}{2}\sqrt{\tilde{V}_{0}\left(1\pm\cos\theta\right)}\frac{\left(x-x_{i}\right)^{2}+\left(y-y_{i}\right)^{2}}{a^{2}}\big]
$
with $\tilde{V}_{0}=\frac{V_{0}}{E_{R}}$.

When expanding the field operator $\psi({\bf r})$ in the basis of the orthonormal Wannier function $w\left({\bf r}-{\bf r}_{i}\right)$, $\psi({\bf r})=\sum\limits_{i}w\left({\bf r}-{\bf r}_{i}\right)\hat{a}_i$, we find that the hopping parameter $t$ and the on-site interaction $U$ are determined by \cite{Jaksch} $
t_{ij}=-\int d{\bf r}w^{*}\left({\bf r}-{\bf r}_{i}\right)\left(-\frac{\hbar^{2}}{2\pi}\nabla^{2}-V_{op}({\bf r})\right)w\left({\bf r}-{\bf r}_{j}\right)
$
and
$
U=g\int d{\bf r}\left|w\left({\bf r}-{\bf r}_{i}\right)\right|^{4}
$,
respectively. Meanwhile, from the expression of the optical lattice potential in Eq. (\ref{lattice-potential-1}), we see that $\Delta \mu=4V_0\cos\theta$.

The density distribution function in momentum space \cite{kashurnikov} reads
$
n\left({\bf k}\right)=\int d{\bf r}d{\bf r}^{\prime}e^{i{\bf k}\cdot\left({\bf r}-{\bf r}^{\prime}\right)}
 \left\langle \psi^{\dagger}\left({\bf r}\right)\psi\left({\bf r}^{\prime}\right)\right\rangle
$
which can in turn be expressed in terms of the ${\hat a}_{\bf k}=\frac1{N_S}\sum\limits_i\hat{a}_ie^{-i{\bf k}\cdot{\bf r}_i}$ as $
n\left({\bf k}\right)=N_{S}\left|w\left({\bf k}\right)\right|^{2}\left\langle \hat{a}_{\bf k}^{\dagger}\hat{a}_{\bf k}\right\rangle
$ \cite{blakie}, or in other words,
$
n\left({\bf k}\right)=N_{S}\left|w\left({\bf k}\right)\right|^{2}\lim\limits_{\tau\downarrow 0}G\left(\tau |0,{\bf k}\right)
$,
where $G\left(\tau |0,{\bf k}\right)\equiv\langle\hat{T}_{\tau}[\hat{a}^{\dagger}(\tau)_{\bf k}\hat{a}(0)_{\bf k}]\rangle$ is the Fourier transformation of the one particle Green's function $G(\tau', j'|\tau, j)\equiv\langle\hat{T}_{\tau}[\hat{a}^{\dagger}_{j'}(\tau')\hat{a}_j(\tau)]\rangle$. By treating the hopping term in Eq.(\ref{H_SL}) as perturbation,  in Dirac picture, the Green's function reads
$
G(\tau^{\prime},j^{\prime}\mid
\tau,j)=\,\frac{\text{Tr}\{{{e^{-\beta
\hat{H}_{0}}}\hat{T}_{\tau}[\hat{a}^{\dagger}_{j^{\prime}}(\tau^{\prime})\hat{a}_{j}(\tau)\hat{u}(\beta,0)]}\}}
{\{\text{Tr}{e^{-\beta \hat{H}_{0}}}\hat{u}(\beta,0)\}}
$,
where
$\hat{O}(\tau)=e^{\tau\hat{H}_{0}}\hat{O}e^{- \tau\hat{H}_{0}}$,
and $\hat{u}(\beta,0)=\hat{T}_{\tau}\left[\exp\left(\int^{\beta}_{0}d\tau\sum_{\langle
i,j \rangle} t
\hat{a}^{\dagger}_{i}(\tau)\hat{a}_{j}(\tau)\right)\right]
$ is the evolution operator (setting
$\hbar=1$) \cite{peskin}. After expanding the evolution operator perturbatively, it is not difficult to see that the expansion of the Green's function consists of terms as
$
\frac{1}{n!}\sum_{i_{1},j_{1},\cdots,i_{n},j_{n}}t_{i_{1},j_{1}}\cdots
t_{i_{n},j_{n}}\int^{\beta}_{0}d\tau_{1} \cdots
\int^{\beta}_{0}d\tau_{n}$ $
\langle\hat{T}_{\tau}[\hat{a}^{\dagger}_{j^{\prime}}(\tau^{\prime})\hat{a}_{j}(\tau)\hat{a}^{\dagger}_{i_{1}}(\tau_{1})
\hat{a}_{j_{1}}(\tau_{1})\cdots\hat{a}^{\dagger}_{i_{n}}(\tau_{n})\hat{a}_{j_{n}}(\tau_{n})]\rangle_{0}
$ with $\langle\;\rangle_0$ being the average quantity with respect to $\hat{H}_0$, in other words, $n$-particle Green's functions  $G_n^{(0)}=\langle\hat{T}_{\tau}[\hat{a}^{\dagger}_{i_{1}}(\tau_{1})
\hat{a}_{j_{1}}(\tau_{1})\cdots\hat{a}^{\dagger}_{i_{n}}(\tau_{n})\hat{a}_{j_{n}}(\tau_{n})]\rangle_{0}$ with respect to $\hat{H}_0$ need to be calculated out, and this can be achieved by expanding them in terms of cumulants \cite{jiang-pra-2,metzner1,ohliger} $C^{(0)}_m(\tau^{\prime}_{1},\cdots,\tau^{\prime}_{m}\mid\tau_{1},\cdots\tau_{m})
=\langle
\hat{T}_{\tau}[\hat{a}^{\dagger}(\tau^{\prime}_{1})\hat{a}(\tau_{1})\cdots
\hat{a}^{\dagger}(\tau^{\prime}_{m})\hat{a}(\tau_{m})]\rangle_{0}
$. However, since we are dealing with a system on a superlattice, according to the generalized Green's function method \cite{lin-arxiv}, the cumulants $C^{(0)}_{mA}=\langle
\hat{T}_{\tau}[\hat{a}^{\dagger}_A(\tau^{\prime}_{1})\hat{a}_A(\tau_{1})\cdots
\hat{a}_A^{\dagger}(\tau^{\prime}_{m})\hat{a}_A(\tau_{m})]\rangle_{0}
$ on sublattice $A$ and cumulants $C^{(0)}_{mB}=\langle
\hat{T}_{\tau}[\hat{a}_B^{\dagger}(\tau^{\prime}_{1})\hat{a}_B(\tau_{1})\cdots
\hat{a}_B^{\dagger}(\tau^{\prime}_{m})\hat{a}_B(\tau_{m})]\rangle_{0}
$ on sublattice $B$ are different and have to be treated separately.

In practice, the Green's functions can only be calculated out perturbatively by selecting specific groups of terms in the above cumulant expansion expressions in a proper way. According to re-summed Green's function method \cite{jiang-pra-2,ohliger,lin-arxiv}, to the lowest order, only terms consisting of first order cumulants are picked up. In terms of Matsubara frequency, the lowest order re-summed one-particle Green's function in momentum space reads $\tilde{G}(\omega_{m},\textbf{k})=\sum_{l=0}^{+\infty}\!C^{(0)}_{1A}(\omega_{m})^{l+1}C^{(0)}_{1B}(\omega_{m})^{l}
\left(t({\bf k})\right)^{2l}
+\sum_{l=0}^{+\infty}C^{(0)}_{1A}(\omega_{m})^{l}C^{(0)}_{1B}(\omega_{m})^{l+1}\left(t({\bf k})\right)^{2l}+2\sum_{l=0}^{+\infty}C^{(0)}_{1A}(\omega_{m})^{l+1}C^{(0)}_{1B}(\omega_{m})^{l+1}\left(t({\bf k})\right)^{2l+1}$
where $C^{(0)}_{1A(B)}(\omega_{m})\!\!=\!\!\int^{\beta}_{0}C^{(0)}_{1A(B)}(\tau)e^{i\omega_{m}\tau} d\tau$ and
$t\left({\bf k}\right)\!\!=\!\!2t\left[\cos\left(k_{x}a\right)\!+\!\cos\left(k_{y}a\right)\right]$. Since $C^{(0)}_{1}\!(\tau)\!=\!\langle\!
\hat{T}_{\tau}[\hat{a}^{\dagger}(\tau) \hat{a}(0)]\!\rangle_{0}$,
by counting the detailed information of the eigenstates of $\hat{H}_0$, in the zero temperature limit, a complicated yet straightforward calculation leads to
$C^{(0)}_{1A}(\omega_{m})=\frac{n_{A}+1}{E_{n_{A}+1,n_{B}}-E_{n_{A},n_{B}}+i\omega_{m}}
-\frac{n_{A}}{E_{n_{A},n_{B}}-E_{n_{A}-1,n_{B}}+i\omega_{m}}$ and
$C^{(0)}_{1B}(\omega_{m})=\frac{n_{B}+1}{E_{n_{A},n_{B}+1}-E_{n_{A},n_{B}}+i\omega_{m}}
-\frac{n_{B}}{E_{n_{A},n_{B}}-E_{n_{A},n_{B}-1}+i\omega_{m}}$
together with the expression of the unperturbed ground state energy $E_{n_A,n_B}$ in Eq.(\ref{eigenvalue}),
the Green's function $G(\tau|0,{\bf k})$ can then be calculated analytically via $
G(\tau|0,{\bf k})=\frac1{2\pi}\int_{-\infty}^{+\infty}d\omega_m \tilde{G}(\omega_m,{\bf k})e^{-\omega_m\tau}$.

However, in order to calculate the one-particle Green's function quantitatively and to plot the corresponding time-of-flight absorption pictures, some basic facts and parameters of possible experiments need to be clarified.
We may propose that the superlattice Bose-Hubbard model may be realized in experiment by trapping ultracold  $^{87}$Rb Bose gas in a cubic optical  lattice created by laser beams with wavelength $\lambda\approx850$ nm, the corresponding $s$-wave scattering length of $^{87}$Rb is estimated to be about $a_{s}\approx5.34$ nm. The superlattice structure is achieved, according to Eq.(\ref{lattice-potential-1}), by tuning the phase difference $\theta$. In order to show the alternate Mott and CDW phases, according to the discussion in section II, we choose $\theta$ properly so that $\Delta \mu/U=0.5$. In experiment, in order to eliminate the influence of the third dimension to reveal the 2D properties of the system, the lattice depth in the third dimension $V_3=30 E_R$ would be a reasonable choice \cite{j,j2,becker}.

\begin{figure}[h!]
\centering
\includegraphics[width=0.24\linewidth]{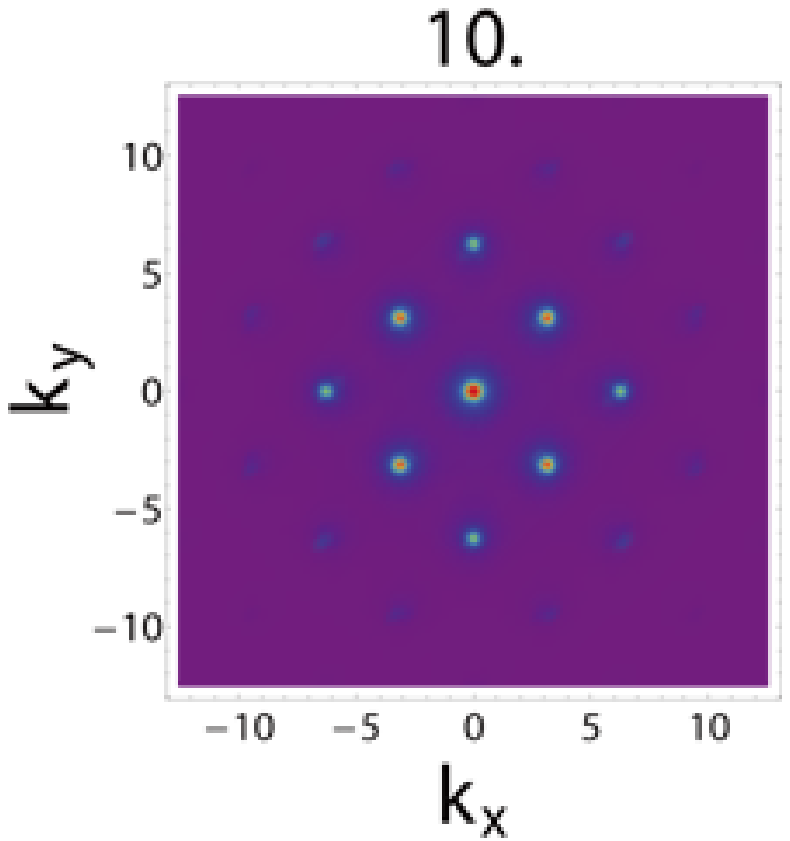}
\includegraphics[width=0.24\linewidth]{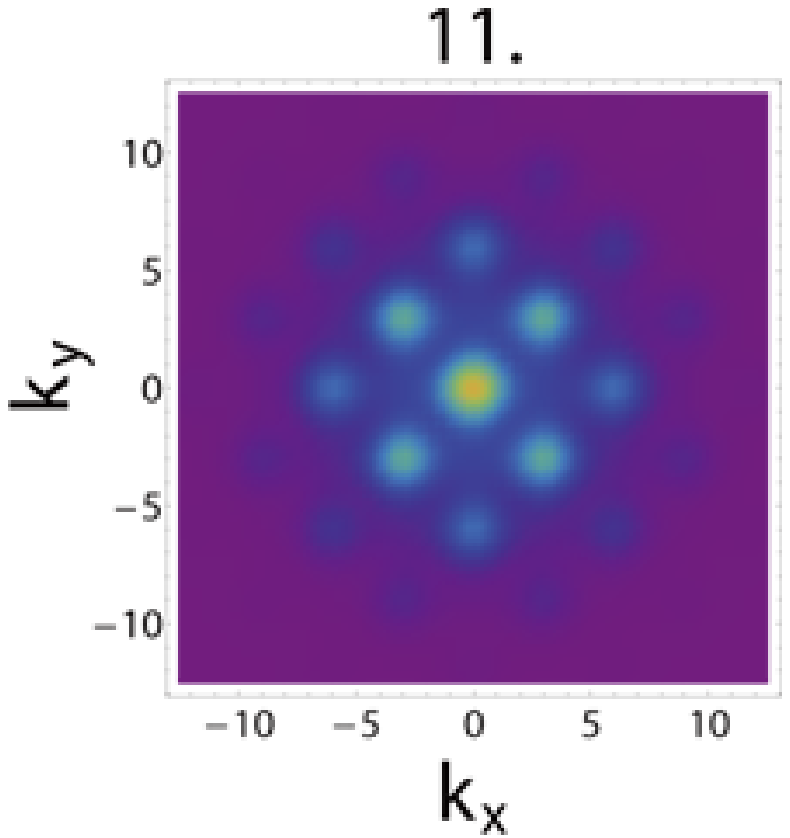}
\includegraphics[width=0.24\linewidth]{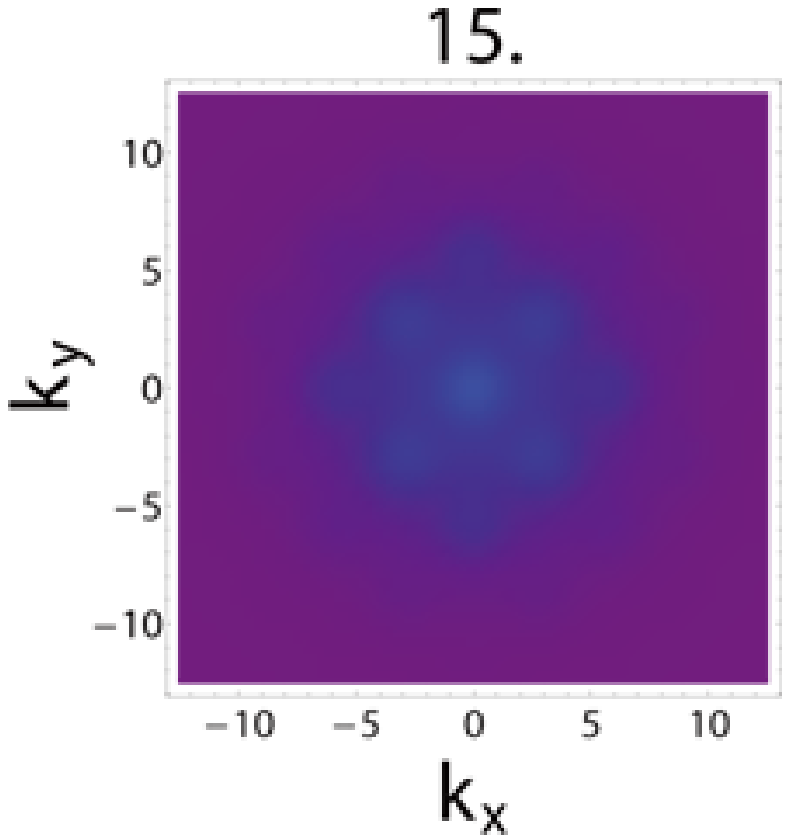}
\includegraphics[width=0.24\linewidth]{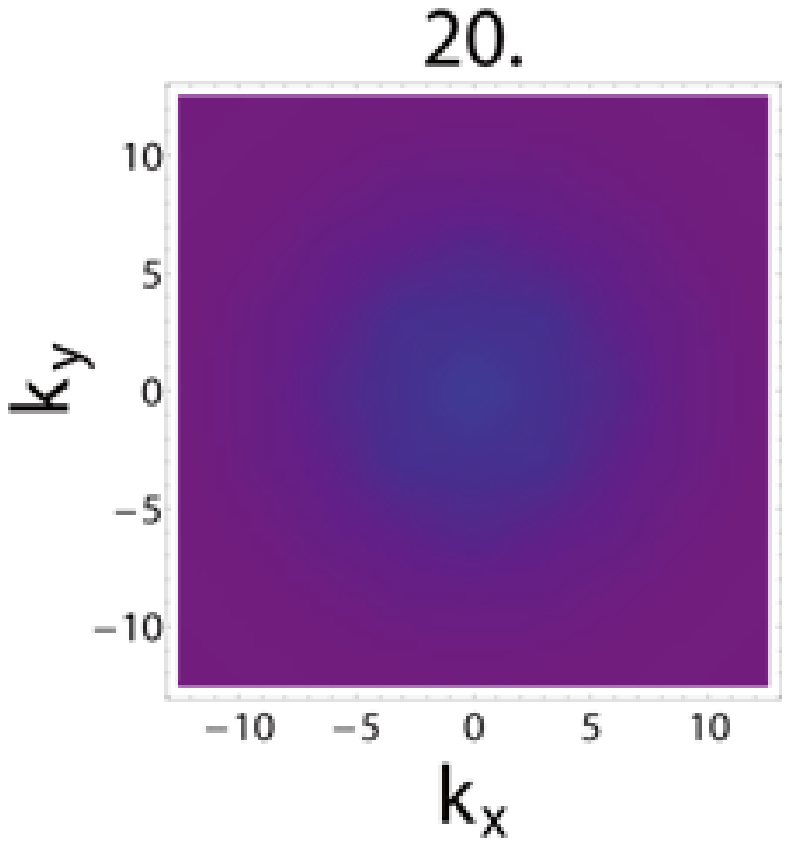}
\caption{The time-of-flight absorption pictures for various $\tilde{V}_0$ for an ultracold Bose gas in a square superlattice with its uncompressed state being CDW(2,1). $k_x$ and $k_y$ in the plots take the unit of $1/a$.}
\label{CDW}
\end{figure}

\begin{figure}[h!]
\begin{center}
\includegraphics[width=0.24\linewidth]{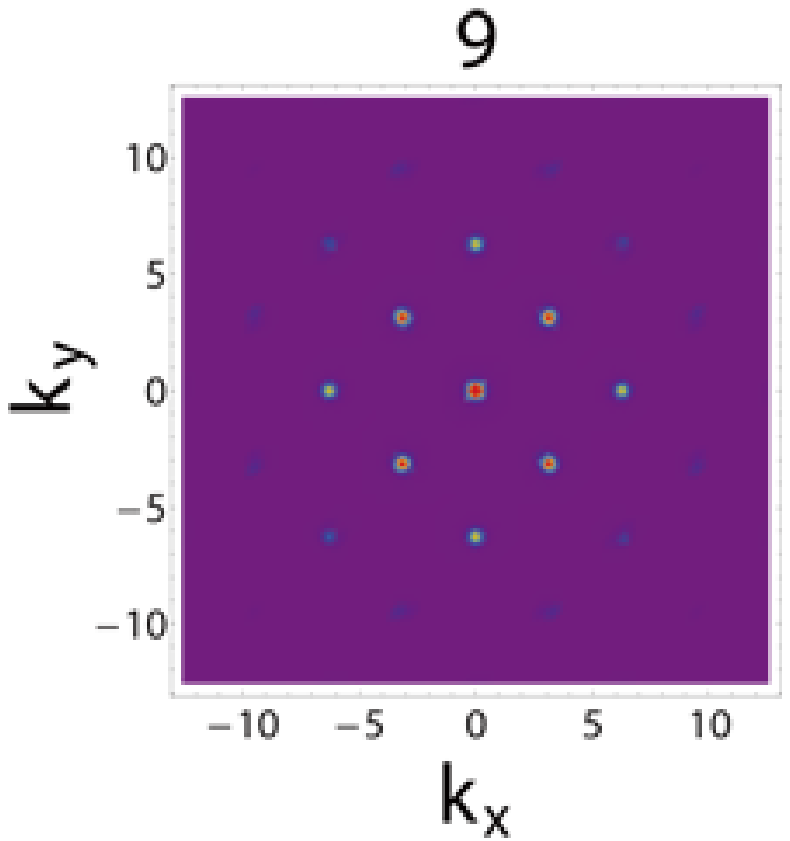}
\includegraphics[width=0.24\linewidth]{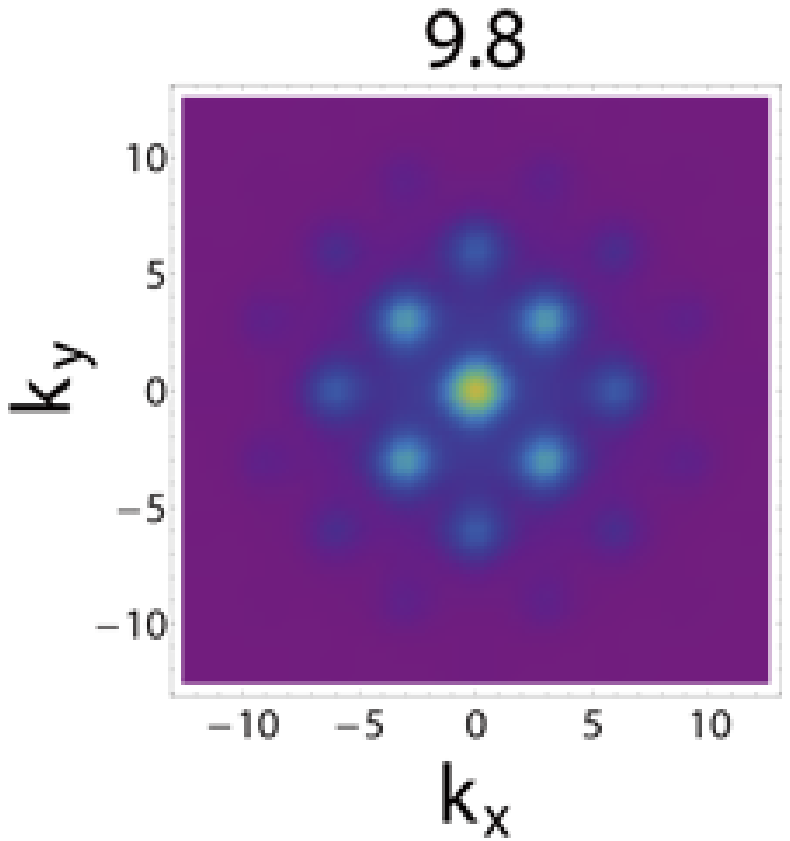}
\includegraphics[width=0.24\linewidth]{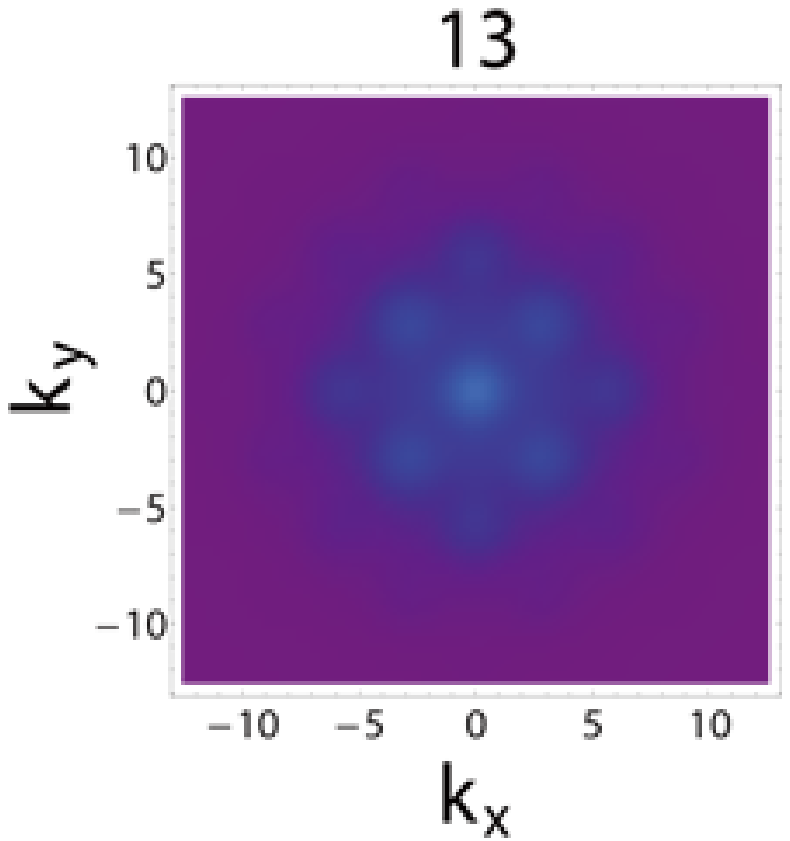}
\includegraphics[width=0.24\linewidth]{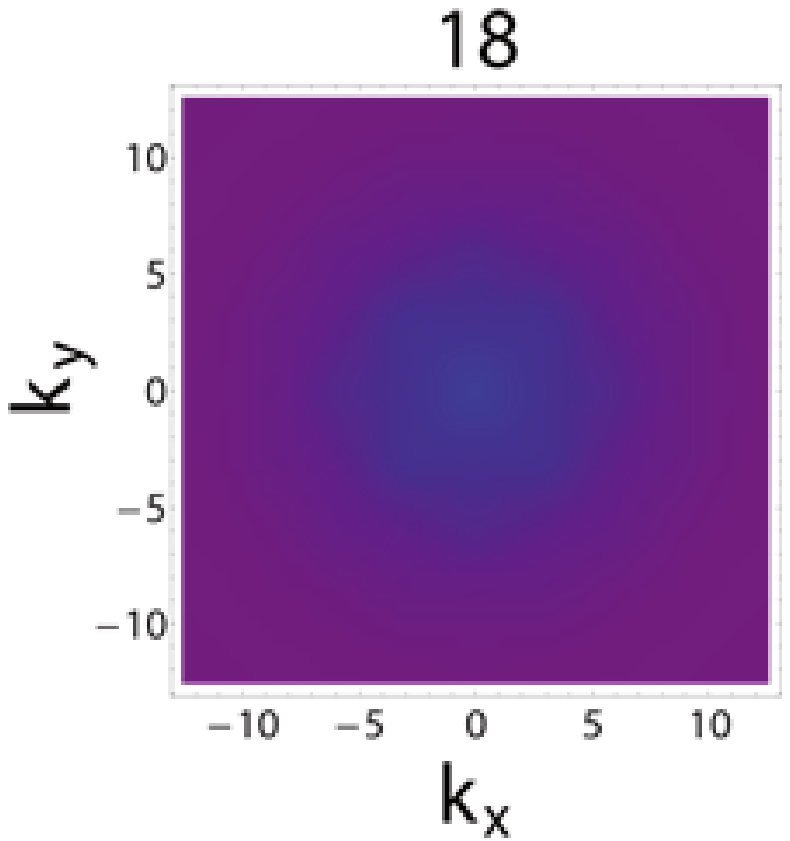}
\caption{The time-of-flight absorption pictures for various $\tilde{V}_0$ for an ultracold Bose gas in a square superlattice with its uncompressed state being MI(1,1). $k_x$ and $k_y$ in the plots take the unit of $1/a$.}
\label{Mot}
\end{center}
\end{figure}

By taking all above information into account, we calculate and plot the time-of-flight absorption pictures of the superlattice ultracold Bose system released from CDW(2,1) state and from MI(1,1) state for different lattice depths in Fig. \ref{CDW} and Fig. \ref{Mot}, respectively, it is clear to see that the phases transit from superfluid phase to Mott or CDW phase when increasing the lattice depth.

From Figs. \ref{CDW} and \ref{Mot}, we see that the quantum phase transitions of CDW(2,1) case start to appear with deeper lattice depth when compared with what in MI(1,1) case, in other words, the quantum critical value of hopping parameter $t_c$ for CDW(2,1) case is smaller than in MI(1,1) case, this is in consistence with the quantum phase diagram in Fig. \ref{phase-boundary}. When compared with the time-of-flight pictures of Bose-Hubbard system in a homogeneous square lattice \cite{Greiner}, there are extra peaks at $(\pm \pi, \pm \pi)$ in present TOF pictures, this is due to the bipartite superlattice structure, and reflects the inhomogeneity of the superfluid phase.

- {\it Acknowledgement.} - The authors greatly acknowledge Axel Pelster and Tao Wang for their stimulating and fruitful
discussions. This Work was supported by National Natural Science Foundation of China under Grant No. 11275119 and by Ph.D. Programs Foundation of Ministry of Education of China under Grant No. 20123108110004.

\end{document}